\documentstyle{article}

\def\bea{\begin{eqnarray}}
\def\eea{\end{eqnarray}}
\def\nn{\nonumber}

\def\beq{\begin{equation}}
\def\eeq{\end{equation}}
\def\ba{\beq\new\begin{array}{c}}
\def\ea{\end{array}\eeq}
\def\be{\ba}
\def\ee{\ea}
\def\stackreb#1#2{\mathrel{\mathop{#2}\limits_{#1}}}

\def\res{{\rm res}}

\makeatletter
\newdimen\normalarrayskip              
\newdimen\minarrayskip                 
\normalarrayskip\baselineskip
\minarrayskip\jot
\newif\ifold             \oldtrue            \def\new{\oldfalse}
\def\arraymode{\ifold\relax\else\displaystyle\fi} 
\def\eqnumphantom{\phantom{(\theequation)}}     
\def\@arrayskip{\ifold\baselineskip\z@\lineskip\z@
     \else
     \baselineskip\minarrayskip\lineskip2\minarrayskip\fi}
\def\@arrayclassz{\ifcase \@lastchclass \@acolampacol \or
\@ampacol \or \or \or \@addamp \or
   \@acolampacol \or \@firstampfalse \@acol \fi
\edef\@preamble{\@preamble
  \ifcase \@chnum
     \hfil$\relax\arraymode\@sharp$\hfil
     \or $\relax\arraymode\@sharp$\hfil
     \or \hfil$\relax\arraymode\@sharp$\fi}}
\def\@array[#1]#2{\setbox\@arstrutbox=\hbox{\vrule
     height\arraystretch \ht\strutbox
     depth\arraystretch \dp\strutbox
     width\z@}\@mkpream{#2}\edef\@preamble{\halign
\noexpand\@halignto
\bgroup \tabskip\z@ \@arstrut \@preamble \tabskip\z@ \cr}%
\let\@startpbox\@@startpbox \let\@endpbox\@@endpbox
  \if #1t\vtop \else \if#1b\vbox \else \vcenter \fi\fi
  \bgroup \let\par\relax
  \let\@sharp##\let\protect\relax
  \@arrayskip\@preamble}
\def\eqnarray{\stepcounter{equation}%
              \let\@currentlabel=\theequation
              \global\@eqnswtrue
              \global\@eqcnt\z@
              \tabskip\@centering
              \let\\=\@eqncr
              $$%
 \halign to \displaywidth\bgroup
    \eqnumphantom\@eqnsel\hskip\@centering
    $\displaystyle \tabskip\z@ {##}$%
    \global\@eqcnt\@ne \hskip 2\arraycolsep
         $\displaystyle\arraymode{##}$\hfil
    \global\@eqcnt\tw@ \hskip 2\arraycolsep
         $\displaystyle\tabskip\z@{##}$\hfil
         \tabskip\@centering
    &{##}\tabskip\z@\cr}
\newfont{\hr}{msbm10}
\newfont{\ams}{msam10}

\begin{document}
\begin{titlepage}
\setcounter{footnote}0
\begin{center}
\hfill FIAN/TD-01/97\\
\hfill ITEP/TH-02/97\\
\hfill hep-th/9701014\\

\vspace{0.3in}
{\LARGE\bf WDVV Equations from Algebra of Forms}
\\ \bigskip\bigskip\bigskip

{\Large A.Marshakov
\footnote{E-mail address:
mars@lpi.ac.ru, andrei@heron.itep.ru, marshakov@nbivms.nbi.dk},
A.Mironov}
\footnote{E-mail address:
mironov@lpi.ac.ru, mironov@heron.itep.ru}\\
{\it Theory Department, P. N. Lebedev Physics
Institute, Leninsky prospect 53, Moscow, ~117924, Russia\\
and ITEP, Moscow ~117259, Russia}\\and\\
{\Large A.Morozov}
\footnote{E-mail address:
morozov@vxdesy.desy.de}
\\
{\it ITEP, Moscow ~117 259, Russia}\\
\end{center}
\bigskip \bigskip

\begin{abstract}
A class of solutions to the WDVV equations is provided by
period matrices of hyperelliptic Riemann surfaces,
with or without punctures. The equations themselves reflect
associativity of explicitly described multiplicative algebra
of (possibly meromorphic) 1-differentials, which holds at
least in the hyperelliptic case. This construction
is direct generalization of the old one, involving the ring
of polynomials factorized over an ideal, and is inspired by
the study of the Seiberg-Witten theory. It has potential to be
further extended to reveal algebraic structures underlying the theory
of quantum cohomologies and the prepotentials in string
models with $N=2$ supersymmetry.
\end{abstract}

\end{titlepage}

\newpage
\setcounter{footnote}0

\section{Introduction}

\subsection{WDVV equations}

The WDVV (Witten-Dijkgraaf-Verlinde-Verlinde) equations \cite{W,DVV,D1}
state that the third derivatives of the {\it prepotential} ${\cal F}(a_i)$
organized in the matrices
\be
({\cal F}_i)_{jk} = {\cal F}_{ijk} = \frac{\partial {\cal F}}
{\partial a_i\partial a_j\partial a_k},
\ee
satisfy \cite{MMM}
\be
{\cal F}_i{\cal F}_k^{-1}{\cal F}_j =
{\cal F}_j{\cal F}_k^{-1}{\cal F}_i \ \ \ \
\forall i,j,k.
\label{WDVV}
\ee
The {\it moduli} $a_i$ are defined up to linear transformations
(i.e. define the {\it flat structure} on the {\it moduli space})
which leave the whole set (\ref{WDVV}) invariant.

The WDVV equations can be reformulated in the following way. Given any
{\it metric}
\be
G = \sum_m g^{(m)}{\cal F}_m
\label{metric}
\ee
one may use it to raise up indices and introduce
\be
C^{(G)}_j = G^{-1}{\cal F}_j,
\label{defC}
\ee
i.e. $C^i_{jk} = (G^{-1})^{im}{\cal F}_{mjk}$, or
${\cal F}_{ijk} = G_{im}C^m_{jk}$. From now on we omit the
superscript $(G)$ in $C^{(G)}$ and assume summation over
repeated indices. Then the WDVV eqs imply that all matrices
$C$ commute:
\be
C_iC_j = C_jC_i \ \ \ \
\forall i,j
\label{comC}
\ee
(and thus can be diagonalized simultaneously).
While (\ref{WDVV}) implies (\ref{comC}), inverse is not true:
the WDVV equations are either (\ref{WDVV}) or the combination of
(\ref{metric}), (\ref{defC}) and (\ref{comC}).

The WDVV eqs were first derived \cite{DVV} in the study of the chiral rings
\cite{chr} in $2d$ $N=2$ superconformal topological models,
where (\ref{comC}) expresses the associativity of the multiplication of
observables $\phi_i$
\be
\phi_i \circ \phi_j = C^k_{ij} \phi_k, \\
(\phi_i\circ\phi_j)\circ\phi_k = \phi_i\circ(\phi_j\circ\phi_k),
\label{alg}
\ee
while
\be
{\cal F}_{ijk} = \langle\langle \phi_i \phi_j \phi_k \rangle\rangle
\ee
are (deformed) 3-point correlation functions on sphere. In this particular
context, there is a distinguished observable $\phi_0 = I$ and associated
distinguished metric
$G^{(0)}_{ij} = \langle\langle \phi_i\phi_j I\rangle\rangle =
{\cal F}_{0ij}$.

\subsection{Polynomial ring \label{Polring}}

The basic example of the algebra (\ref{alg}) is the multiplication
of polynomials modulo $dP$ (in the string-theory language this is the case of
the {\it Landau-Ginzburg} topological models):  \be
\phi_i(\lambda)\phi_j(\lambda) = C^k_{ij} \phi_k(\lambda)G'(\lambda)
\ {\rm mod}\ P'(\lambda)
\ee
Here $P(\lambda)$ and $G(\lambda)$
are polynomials of $\lambda$,
such that their $\lambda$-derivatives $P'(\lambda)$
and $G'(\lambda)$  are co-prime (do not have common divisors),
and $\phi_i(\lambda)$ form a complete basis in the linear space of polynomials
modulo $P'(\lambda)$. Thus it gives a particular case of the algebra
(\ref{alg})
\be
\phi_i(\lambda)\circ \phi_j(\lambda) = C_{ij}^k\phi_k(\lambda),
\label{algp}
\ee
and it is associative as a factor of explicitly
associative multiplication algebra of polynomials over its ideal
$P'(\lambda) = 0$. The structure constants $C_{ij}^k$ depend
on the choice of $P(\lambda)$ (the point of the
"moduli" space) and $G'(\lambda)$ (the metric).

The second ingredient of the WDVV eqs is the residue formula \cite{ref},
\be
{\cal F}_{ijk} = \stackreb{dP = 0}{\res}\
\frac{\phi_i(\lambda)\phi_j(\lambda)\phi_k(\lambda)}{P'(\lambda)}
d\lambda
\label{resf}
\ee
In accordance with (\ref{defC}),
\be
G'(\lambda) = g^{(m)}\phi_m(\lambda)\ \ \ \ \ \ \
G_{ij} = g^{(m)}{\cal F}_{ijm}
\ee
The last ingredient is the expression of {\it flat} moduli
$a_i$ in terms of the polynomial $P(\lambda)$ \cite{KriWhi}:
\be
a_i = -{N\over i(N-i)}\res \left( P^{i\over N}dG \right),\ \
N = {\rm ord}(P)
\ee
These formulas have a straightforward generalization to the
case of polynomials of several variables, $\phi_i(\vec\lambda)
= \phi_i(\lambda_1,\ldots,\lambda_n)$:
\be
\phi_i(\vec\lambda)\phi_j(\vec\lambda) =
C_{ij}^k\phi_k(\vec\lambda)Q(\vec\lambda) \ {\rm mod} \ \left(
\frac{\partial P}{\partial\lambda_1},\ldots,\frac{\partial P}
{\partial \lambda_n}\right),
\ee
and
\be
{\cal F}_{ijk} = \stackreb{dP = 0}{\res}\
\frac{\phi_i(\vec\lambda)\phi_j(\vec\lambda)\phi_k(\vec\lambda)}
{\prod_{\alpha=1}^n \frac{\partial P}{\partial \lambda_\alpha}}
d\lambda_1\ldots d\lambda_n
\label{resf1}
\ee
The algebra (\ref{algp}) is always associative, since
$dP = \sum_{\alpha=1}^n \frac{\partial P}{\partial \lambda_\alpha}
d\lambda_\alpha$ is always an ideal in the space of polynomials.
Moreover, one can even take a factor over generic ideal in the
space of polynomials, $p_1(\vec\lambda) = \ldots = p_n(\vec\lambda) = 0$,
where polynomials $p_\alpha$ need to be co-prime, but do not need to
be derivatives of a single $P(\vec\lambda)$.

\subsection{Other examples}

The above example of the polynomial ring is very transparent since it is related
to an obviously associative algebra of polynomials, and associativity is
preserved by factorization over an ideal. Less transparent are the origins of
the residue formula and expression for the moduli $a_i$. This problem
is, however, resolved in a general framework \cite{MMM,MMM2},
inspired by the Seiberg-Witten theory \cite{SW}.

Before turning to the general situation one should mention that the main stream
of study of the WDVV eqs has been so far in another direction.  One of the
most interesting questions is related to
the deformations of the polynomial ring, associated with the
Gromov-Witten (GW) classes \cite{W,KoMa}, or quantum cohomologies \cite{M}.
For the rational (coming from {\it rational} curves) GW classes, the WDVV eqs
(\ref{WDVV}) are still
true, but no nice description in terms of ideals of the obviously
associative algebra is known yet (or, better, no nice way is yet known to
specify the moduli dependence of the ideals). To prove the equations
without explicit associative algebra, the sophisticated methods were
developed, relating them to the theory of the Frobenius algebras and Egoroff
metrics \cite{D1}, and to the properties of the moduli spaces ${\cal M}_{0,n}$
\cite{KoMa,M}.  Appropriate generalizations of the WDVV eqs to higher genus
GW classes and to higher dimensions (from world-sheet instantons inspired by
strings to world-volume ones inspired by branes) are difficult to find in
such a framework (see, however, \cite{P} and \cite{G} for some results about
elliptic case).

Recently, the WDVV equations appeared in a naively different
context \cite{MMM} (see also \cite{BM} and \cite{BW}):
as equations on the prepotentials in the Seiberg-Witten theory
\cite{SW}, describing the low-energy limit of $N=2$ supersymmetric Yang-Mills
models in $4d$. Remarkably, the proof of these equations, suggested
in \cite{MMM,MMM2}, appeared to be actually a return to the approach
used in sect.\ref{Polring}: the equations are related to an obviously
associative multiplication algebra. What happens is that the polynomials
(functions on a Riemann sphere) are substituted by the holomorphic
1-differentials on Riemann surfaces (complex curves). They always form
a family of closed algebras, parametrized by a triple of
holomorphic differentials $dG,dW,d\Lambda$. However, these algebras
are not rings in the usual sense of the word, thus they are not immediately
associative after factorization over an ideal. Still, associativity
is preserved for the classes of hyperelliptic curves
appearing in the Seiberg-Witten theory \cite{GKMMM}.

The purpose of this letter is to give a brief presentation of this
construction.

Clearly, it should possess direct generalizations to higher complex
dimensions (from holomorphic 1-forms on complex curves to forms on
complex manifolds), one can even think that it would provide a new
look at the theory of quantum cohomologies. The very fact that the WDVV
equations hold for the
Seiberg-Witten prepotentials can also imply that there exist universal
equations for the prepotentials in string models (which turn into the
Seiberg-Witten prepotentials in
certain limiting cases). Such equations are not yet known for generic
families of Calabi-Yau spaces (the WDVV eqs, inspired by the theory of
quantum cohomologies, are empty for Calabi-Yau threefolds).
The Picard-Fuchs equations, which are always true, are {\it not
universal} -- they depend strongly on peculiarities of particular
family. We shall deliberately ignore further comments on these possible
generalizations in this letter, and concentrate on the case of
the complex curves.

The construction itself is described in the next section 2, and section 3
lists particular examples
(families of hyperelliptic
curves and triples $dG$, $dW$, $d\Lambda$), which have been analyzed in this
framework. Further technical details about these examples can be
found in \cite{MMM2}.

\section{WDVV equations for the families of hyperelliptic curves}

\subsection{WDVV eqs from associativity and residue formula
\label{basic}}

As we already mentioned, the WDVV equations can be considered as
a synthesis of the two ingredients: associativity of algebra
and residue formula for prepotential. Namely, imagine that
in some context the following statements are true:
\begin{enumerate}
\item The holomorphic
\footnote{Since curves with punctures and the corresponding
meromorphic differentials
can be obtained by degeneration of smooth curves of higher genera we do not
make any distinction between punctured and smooth curves below. We remind that
the holomorphic 1-differentials can have at most simple poles at the punctures
while quadratic differentials can have certain double poles etc.}
1-differentials
on the complex curve ${\cal C}$ of genus $g$ form a {\it closed} algebra,
\be
d\omega_i(\lambda)d\omega_j(\lambda) = C_{ij}^kd\omega_k(\lambda)
dG(\lambda) + D_{ij}^kd\omega_k(\lambda)dW(\lambda)
+ E_{ij}^kd\omega_k(\lambda)d\Lambda(\lambda) = \nn \\ =
C_{ij}^kd\omega_k(\lambda)dG(\lambda) \ {\rm mod} \ (dW,d\Lambda),
\label{algdiff}
\ee
where $d\omega_i(\lambda)$, $i=1,\ldots, g$, form a complete basis in the
linear space $\Omega^1$ (of holomorphic 1-forms),
$dG$, $dW$ and $d\Lambda$ are fixed elements of $\Omega^1$, e.g.
$dG(\lambda) = \sum_{m=1}^g  \eta^{(m)}d\omega_m$.
\item The factor of this algebra over the ``ideal'' $dW\oplus d\Lambda$
is associative,
\be
C_i C_j = C_jC_i \ \ \forall  i,j \ {\rm at\ fixed}\
dG, dW, d\Lambda
\label{comCdiff}
\ee
(remind that $(C_i)^k_j \equiv C_{ij}^k$).
\item The residue formula holds,
\be
\frac{\partial {\cal F}}{\partial a_i\partial_j\partial a_k}
= \stackreb{dW = 0}{\res } \frac{d\omega_id\omega_jd\omega_k}
{dWd\Lambda} =
-\stackreb{d\Lambda = 0}{\res} \frac{d\omega_id\omega_jd\omega_k}
{dWd\Lambda}
\label{resfdiff}
\ee
\item There exists a non-degenerate linear combination of matrices
${\cal F}_i$.
\end{enumerate}
These statements imply the WDVV eqs (\ref{WDVV}) for the prepotential
${\cal F}(a_i)$ \cite{MMM}. Indeed, the
substitution of (\ref{algdiff}) into (\ref{resfdiff}) gives
\be
{\cal F}_{ijk} = C^m_{ij}G_{mk},
\label{int1}
\ee
where
\be
G_{mk} = \stackreb{dW = 0}{\res} \frac{dG d\omega_md\omega_k}
{dWd\Lambda} = \eta^{(l)}{\cal F}_{lmk},
\ee
and the terms with $dW$ and $d\Lambda$ in (\ref{algdiff}) drop
out from ${\cal F}_{ijk}$ because they cancel $d\Lambda$ or $dW$
in the denominator in (\ref{resfdiff}).
Eq.(\ref{int1}) can be now substituted into (\ref{comCdiff}) to
provide WDVV eqs in the form
\be
{\cal F}_i G^{-1} {\cal F}_j = {\cal F}_j G^{-1} {\cal F}_i,
\ \ \  G =  \eta^{(m)}{\cal F}_m \ \ \ \forall\  \left\{\eta^{(m)}\right\}
\label{WDVVdiff}
\ee
where at least one invertible metric $G$ exists by requirement (4).

Actually (\ref{WDVVdiff}) for all (non-degenerate) $G$ follows
immediately from that for some particular $G = \hat G$.\footnote{We
are indebted to A.Rosly for this simple argument.}
Indeed, if all the $\hat C_i = \hat G^{-1} {\cal F}_i$ mutually commute,
then  $G =  \eta^{(m)}{\cal F}_m = \hat G \eta^{(m)} \hat C_m$ and
\be
{\cal F}_iG^{-1}{\cal F}_j =
\hat G \left(\hat C_i \left(\eta^{(m)} \hat C_m\right)^{-1}\hat C_j\right)
\nn
\ee
is obviously symmetric under permutation $i\leftrightarrow j$
(because of commutativity of matrices $\hat C$'s).

Thus, we see that the real issue in the study of WDVV eqs is
to reveal when the conditions (1)-(4) are true.

\subsection{Algebra of holomorphic (1,0)-forms \label{alge}}

Existence of the multiplication algebra (\ref{algdiff}) is a rather
general feature of compact complex manifolds. Indeed, there are $g$
holomorphic 1-differentials on the complex curve of genus $g$.
However, their products $d\omega_id\omega_j$ are not linearly
independent: they belong to the $3g-3$-dimensional space
$\Omega^2$ of the holomorphic quadratic differentials. Given
three holomorphic 1-differentials $dG$, $dW$, $d\Lambda$, one can
make an identification
\be
\Omega^1\cdot\Omega^1 \in \Omega^2 \cong \Omega^1\cdot
(dG \oplus dW \oplus d\Lambda)
\ee
which in particular basis is exactly (\ref{algdiff}).
For given $i,j$ there are $3g$ adjustment parameters $C_{ij}^k$,
$D_{ij}^k$ and $E_{ij}^k$ at the r.h.s. of (\ref{algdiff}), with
3 "zero modes" -- in the directions $dGdW$, $dGd\Lambda$
and $dWd\Lambda$ (i.e. one can add $dW$ to $C_{ij}^kd\omega_k$ and
simultaneously subtract $dG$ from $D_{ij}^kd\omega_k$). Thus we get
exactly $3g-3$ parameters to match the l.h.s. of
(\ref{algdiff}) -- this makes decomposition (\ref{algdiff}) existing
and unique.

\subsection{Associativity}

Thus we found that the existence of the {\it closed} algebra
(\ref{algdiff}) is a general feature, in particular it does not make any
restrictions on the choice of Riemann surfaces.
However, this algebra is not a ring: it maps the square of
$\Omega^1$ into {\it another} space: $\Omega^1\otimes\Omega^1 \rightarrow
\Omega^2 \neq \Omega^1$. Thus, its factor over the condition
$dW = d\Lambda = 0$ is not guaranteed to have all properties of the ring.
In particular the factor-algebra
\be
d\omega_i\circ d\omega_j = C_{ij}^kd\omega_k
\label{facalgdiff}
\ee
does not need to be associative, i.e. the matrices
$C$ alone (neglecting $D$ and $E$) do not necessarily commute.

However, the associativity would follow if the expansion of $\Omega^3$
(the space of the holomorphic 3-differentials containing
the result of triple multiplication $\Omega^1\cdot\Omega^1\cdot\Omega^1$),
\be
\Omega^3 = \Omega^1\cdot dG\cdot dG \oplus \Omega^2\cdot dW \oplus
\Omega^2\cdot d\Lambda
\label{ass}
\ee
is unique. Then it is obvious that
\be
0 = (d\omega_id\omega_j)d\omega_k - d\omega_i(d\omega_jd\omega_k) =
\left( C_{ij}^lC_{lk}^m - C_{il}^mC_{jk}^l\right)d\omega_m dG^2
\ {\rm mod} (dW,d\Lambda)
\label{ass2}
\ee
would imply $[C_i,C_k] = 0$. However, the dimension of
$\Omega^3$ is $5g-5$, while the number of adjustment parameters
at the r.h.s. of (\ref{ass}) is $g + 2(3g-3) = 7g-6$, modulo only
$g+2$ zero modes (lying in $\Omega^1\cdot dWd\Lambda$,
$\Omega^1\cdot dWdG^2$ and $\Omega^1\cdot d\Lambda dG^2$). For $g>3$ there
is no match:
$5g-5 < 6g-8$, the expansion (\ref{ass}) is not unique, and
associativity can (and does)
\footnote{See \cite{MMM2} for an explicit example of {\it non}-associativity
(actually, this happens in the important Calogero model).}
break down unless there is some special reason for it to survive.

This special reason can exist if the curve ${\cal C}$ has specific symmetries.
The most important example is the set of curves
with an involution $\sigma:\ {\cal C} \rightarrow {\cal C}$,
$\sigma^2=1$, such that all
$\sigma(d\omega_i) = -d\omega_i$, while $\sigma(dW) = -dW$,
$\sigma(d\Lambda) = +d\Lambda$. To have $d\Lambda$ different from
all $d\omega_i$ one should actually take it away from $\Omega^1$,
e.g. allow it to be meromorphic. This also requires some reexamination
of our reasoning in the sect.\ref{alge} we are now going to turn to.

\subsection{Associative algebra of holomorphic 1-forms on hyperelliptic
surfaces}

The hyperelliptic curves are described by the equation
\be
Y^2 = {\rm Pol}_{2g+2}(\lambda),
\ee
and the involution is $\sigma: (Y,\lambda )
\rightarrow (-Y, \lambda )$. The space of holomorphic differentials
is $\Omega^1 = {\rm Span}\left\{\frac{\lambda^\alpha d\lambda}{Y(\lambda)}
\right\}$,
$\alpha = 0,\ldots,g-1$.  This space is odd under $\sigma$,
$\sigma(\Omega^1) = -\Omega^1$, and an example of the (meromorphic)
1-differential which is {\it even} is
\be
d\Lambda = \lambda^rd\lambda,
\ee
$\sigma(d\Lambda) = +d\Lambda$. We will assume that $dG$ and $dW$ still belong
to $\Omega^1$ and thus are $\sigma$-odd. In the case of hyperelliptic
curves with punctures, $\Omega^1$ can include also $\sigma$-even
holomorphic 1-differentials (like $\frac{d\lambda}{(\lambda - \alpha_1)
(\lambda - \alpha_2)}$ or just $d\Lambda$), in such cases we consider
the algebra (\ref{algdiff}) of the $\sigma$-odd holomorphic differentials
$\Omega^1_-$, and assume that $d\omega_i$, $dG$ and $dW$ belong to
$\Omega^1_-$, while $d\Lambda \in \Omega^1_+$.

The spaces $\Omega^2$ and $\Omega^3$ also split into $\sigma$-even
and $\sigma$-odd parts: $\Omega^2 = \Omega^2_+ \oplus \Omega^2_-$
and $\Omega^3 = \Omega^3_+ \oplus \Omega^3_-$. Multiplication
algebra maps $\Omega^1_-$ into $\Omega^2_+$ and further into
$\Omega^3_-$, which have dimensions  $2g-1+2n$ and $3g-2+3n$
respectively. Here $n$ enumerates the punctures, where holomorphic
1-differentials are allowed to have simple poles, while quadratic
and the cubic ones
have at most second- and third-order poles respectively. For our
purposes we assume that punctures on the hyperelliptic curves
enter in pairs: every
puncture is accompanied by its $\sigma$-image.
Parameter $n$ is the number of these {\it pairs},
and the dimension of $\Omega^1_-$ is $g+n$.

Obviously, if all the $d\omega_i$ in (\ref{algdiff})
are from $\Omega^1_-$, then all $E^k_{ij} = 0$, i.e. we actually
deal with the decomposition
\be
\Omega^2_+ = \Omega^1_-\cdot dG + \Omega^1_-\cdot dW
\label{2dec}
\ee
Parameter count now gives:
$2g-1 + 2n = 2(g+n) - 1$ where $-1$ is for the zero mode
$dGdW$. Thus, the hyperelliptic reduction of the algebra (\ref{algdiff})
does exist.

Moreover, it is associative, as follows from consideration
of the decomposition
\be
\Omega^3_- = \Omega^1_-\cdot dG^2 + \Omega^2_+ \cdot dW
\label{3dec}
\ee
Of crucial importance is that now there is no need to include
$d\Lambda$ in this decomposition, since it does not appear at the
r.h.s. of the algebra itself.
Parameter count is now:
$3g-2 + 3n = (g+n) + (2g-1+ 2n) -1$ (there is the unique zero mode
$dWdG^2$). Thus, we see that this time decomposition (\ref{2dec})
is unique, and our algebra is indeed associative.

In fact, one could come to the same conclusions much easier just
noting that all elements of $\Omega^1_-$ are of the form
\be
d\omega_i = \frac{\phi_i(\lambda) d\lambda}{YQ(\lambda)},
\ee
where all $\phi_i(\lambda)$ are polynomials and $Q(\lambda)
= \prod_{\iota =1}^n(\lambda - m_{\iota})$ is
some new polynomial, which takes into account the possible
singularities at punctures $\left(m_{\iota },\pm Y(m_{\iota })\right)$.
Then our algebra is just the one of the polynomials $\phi_i(\lambda)$
and it is existing and associative just for the reasons discussed
in sect.\ref{Polring}. The reasoning in this section can be easily
modified in the case when hyperelliptic curve possesses an extra involution.
The families of such curves appear in the Seiberg-Witten context for the
groups $SO(N)$ and $Sp(N)$: the extra involution in these cases is
$\rho :\lambda\rightarrow -\lambda$. Then one considers $\Omega^1_{--}$
instead of just $\Omega^1_-$ (see \cite{MMM2} for further details).

\subsection{Residue formula}

Let us now forget for a while about the hyperelliptic curves and
discuss the most general {\it raison d'etre} of the residue formula.
It is essentially implied \cite{MMM2} by the Seiberg-Witten theory \cite{SW},
the Dubrovin-Krichever-Novikov theory of Whitham hierarchies
\cite{KriWhi,Whith} and Hitchin's description of integrable models \cite{Hit}.

Namely, imagine that we consider an integrable model
with a Lax operator ${\cal L}(w)$, which is a $N\times N$
matrix-valued function
on a {\it bare} spectral curve $E$, $w \in E$, which is
usually torus or sphere.
Then one can introduce a family of complex curves,
defined by the spectral equation
\be
{\cal C}:\ \ \ \det({\cal L}(w)-\lambda) = 0
\ee
The family is parametrized by the {\it moduli} that in this
context are values of the $N$ Hamiltonians of the system
(since Hamiltonians commute with each other, these are actually
$c$-numbers). We obtain this family in a peculiar parametrization,
which represents the {\it full} spectral curves ${\cal C}$
as the ramified $N$-sheet coverings  over the  {\it bare} curve
$E$,
\be
{\cal P}(\lambda; w) = 0,
\label{curveq}
\ee
where ${\cal P}$ is a polynomial of degree $N$ in $\lambda$.

The fact that we started from a Hamiltonian (integrable) system provides us
with additional structure: the symplectic form on the ``bundle'' ${\cal C}
\rightarrow {\cal M}$ (${\cal M}$ is the moduli space).
It defines a ``generating'' form
$dS = \Lambda dW$ on every ${\cal C}$, which possesses the
property:
\be
\frac{\partial dS}{\partial {\rm moduli}} \in \Omega^1,
\label{holreq}
\ee
i.e. every variation of $dS$ with the change of moduli is
a holomorphic differential on ${\cal C}$ (normally, even if
differential is holomorphic, its moduli-derivative is not).

This structure allows one to define the holomorphic differentials
in a rather explicit form. Let $s_I$ denote some coordinates on the
moduli space ${\cal M}$. Then
\be
\frac{\partial dS}{\partial s_I} \cong \frac{\partial\Lambda}
{\partial s_I} dW = -\frac{\partial {\cal P}}{\partial s_I}
\frac{dW}{{\cal P}'} \equiv dv_I,
\label{defdv}
\ee
and $dv_I$ provide a set of holomorphic differentials on ${\cal C}$.
It is easy to see that they are indeed holomorphic -- the
variation of (\ref{curveq}) at constant moduli gives:
\be
{\cal P}'d\Lambda + \frac{\partial{\cal P}}{\partial w} dw = 0,
\ee
i.e. the zeroes of ${\cal P}'$ are always the zeroes of $dw$.
Note that prime denotes the derivative with respect to $\Lambda$,
which can be different from the $\lambda$-derivatives.

The set of $dv_I$ is not necessarily the same as $\Omega^1_-$,
it can be both smaller and bigger (in the latter case some $dv_I$ are linearly
dependent). It is a special requirement (standard in the context of
integrable theories) that the family (\ref{curveq}) and generating
differential $dS$ give rise to $dv_I$'s forming a complete
basis in $\Omega^1_-$ (or in $\Omega^1_{--}$). The finite-gap and Hitchin-like
integrable systems provide a large class of examples when this is true.

The prepotential ${\cal F}(a_I)$ for the Hamiltonian system is defined in
terms of the cohomological class of $dS$:
\be
a_I = \oint_{A_I} dS, \nn \\
\frac{\partial {\cal F}}{\partial a_I} = \oint_{B_I} dS, \nn \\
A_I \odot B_J = \delta_{IJ}
\label{defprep}
\ee
The cycles $A_I$
include the $A_i$'s wrapping around the handles
of ${\cal C}$ and $A_\iota$'s going around the punctures.
The conjugate contours $B_I$ include the cycles $B_i$ and the
{\it non-closed} contours $B_\iota$ ending in the singularities
of $dS$ (see sect.5 of \cite{IM2} for more details).

The self-consistency of the definition (\ref{defprep}) of ${\cal F}$,
i.e. the symmetricity of the {\it period matrix}
$\frac{\partial^2 F}{\partial a_I\partial a_J}$ is guaranteed
by the following reasoning.
Let us differentiate equations (\ref{defprep}) with respect to
moduli $s_K$ and use (\ref{defdv}). Then we get:
\be
\int_{B_I} dv_K = \sum_J T_{IJ}\oint_{A_J} dv_K.
\ee
where the second derivative
\be
\frac{\partial^2 F}{\partial a_I\partial a_J} = T_{IJ}
\ee
is the period matrix of the (punctured) Riemann surface
${\cal C}$.
As any period matrix, it is symmetric
\be
\sum_{IJ} (T_{IJ} - T_{JI}) \oint_{A_I}dv_K\oint_{A_J}dv_L =
\sum_I\left(\oint_{A_I}dv_K\int_{B_I}dv_L -
\int_{B_I}dv_K\oint_{A_I}dv_L\right) =  \nn \\ =
{\res}\left( v_K dv_L \right) = 0
\ee
Note also that the holomorphic differentials
$dv_I$, associated with the {\it flat} moduli $a_I$ are {\it canonical}
$d\omega_I$ such that
$\oint_{A_I}d\omega_J = \delta_{IJ}$
and $\oint_{B_I}d\omega_J = T_{IJ}$.

In order to derive the residue formula one should now consider
the moduli derivatives of the period matrix.
It is easy to get:
\be
\sum_{IJ} \frac{\partial T_{IJ}}{\partial s_M}
\oint_{A_I}dv_K\oint_{A_J}dv_L =
\sum_I\left(\oint_{A_I}dv_K\int_{B_I}\frac{\partial dv_L}{\partial s_M} -
\int_{B_I}dv_K\oint_{A_I}\frac{\partial dv_L}{\partial s_M}\right) =
\nn \\ =
\res \left( v_K \frac{\partial dv_L}{\partial s_M}\right)
\label{prom1}
\ee
The r.h.s. is non-vanishing, since differentiation w.r.t. moduli
produces new singularities. From (\ref{defdv})
\be
-\frac{\partial dv_L}{\partial s_M} =
\frac{\partial^2{\cal P}}{\partial s_L\partial s_M}
\frac{dW}{{\cal P}'} +
\left(\frac{\partial{\cal P}}{\partial s_L}\right)'
\left(-\frac{\partial{\cal P}}{\partial s_M}\right)
\frac{dW}{{\cal P}'} -
\frac{\partial{\cal P}}{\partial s_L}
\frac{\partial{\cal P}'}{\partial s_M}
\frac{dW}{({\cal P}')^2} +
\frac{\partial{\cal P}}{\partial s_L}
\frac{\partial{\cal P}}{\partial s_M}
{{\cal P}''dW\over ({\cal P}')^3}
= \nn \\ =
\left[
\left(\frac{ {\partial{\cal P}/\partial s_L}
 {\partial{\cal P}/\partial s_M} }{ {\cal P}'}  \right)' +
\frac{\partial^2{\cal P}}{\partial s_L\partial s_M} \right]
\frac{dW}{{\cal P}'}
\ee
and new singularities (second order poles)
are at zeroes of ${\cal P}'$ (i.e. at those of $dW$). Note
that the contributions from the singularities of
$\partial{\cal P}/\partial s_L$, if any, are already taken into
account in the l.h.s. of (\ref{prom1}).
Picking up the coefficient at the leading singularity, we obtain:
\be
{\res} v_K \frac{\partial dv_L}{\partial s_M} = -
\stackreb{dW = 0}{\res}
\frac{\partial{\cal P}}{\partial s_K}
\frac{\partial{\cal P}}{\partial s_L}
\frac{\partial{\cal P}}{\partial s_M}
\frac{dW^2}{({\cal P}')^3 d\Lambda} =
\stackreb{dW = 0}{\res}
\frac{dv_Kdv_Ldv_M}{dWd\Lambda}
\ee
The integrals at the l.h.s. of (\ref{prom1}) serve to convert
the differentials $dv_I$ into canonical $d\omega _I$. The same matrix
$\oint_{A_I} dv_J$ relates the derivative w.r.t. the moduli $s_I$
and the periods $a_I$. Putting all together we obtain
(see also \cite{Whith}):
\be
\frac{\partial T_{IJ}}{\partial s^K} =
\stackreb{dW = 0}{\res}
\frac{d\omega_Id\omega_Jdv_K}{dW d\Lambda} \nn \\
\frac{\partial^3 F}{\partial a^I\partial a^J\partial a^K} =
\frac{\partial T_{IJ}}{\partial a^K} =
\stackreb{dW = 0}{\res}
\frac{d\omega_Id\omega_Jd\omega_K}{dW d\Lambda}
\label{resfor}
\ee
Note that these formulas essentially depend only on the symplectic
structure $dW \wedge d\Lambda$: e.g. if one makes an infinitesimal shift of
$dW$ by $d\Lambda$, then $(dWd\Lambda)^{-1}$ is shifted by
$-(dW)^{-2}$, i.e. the shift does not contain poles at $d\Lambda = 0$
and thus does not contribute to the residue formula.

\subsection{Summary}

We described the rather general origins of associative algebra of
holomorphic 1-forms and residue formulas. We saw that associativity
requires restriction to particular families of the complex curves,
for example, hyperelliptic ones. In their turn, residue formulas need
the Hitchin-like families of curves, which are peculiar ramified
coverings.
All these requirements are satisfied simultaneously for the
families of hyperelliptic curves, associated with certain
integrable systems: exactly the ones relevant for most examples
discussed in the Seiberg-Witten theory. Some of them will be mentioned
in the next sect.\ref{Exa}.

The main example, when the WDVV eqs are {\it not} true \cite{MMM2},
is the elliptic (Calogero) case, when the {\it bare} spectral curve
is elliptic (so that ${\cal C}$
is no longer hyperelliptic). In this case the WDVV eqs require
substantial modifications, which are yet to be discovered (presumably,
it can be related to the theory of elliptic Gromov-Witten classes).
One of the things that happens in such examples, is the violation of
condition (4) from sect.\ref{basic}: if all the moduli are
taken into account, all the possible metrics $G$ become degenerate
(as the corollary of conformal invariance of the $4d$ model).
Instead, new non-trivial moduli enter the game, like the parameter
$\tau$ of elliptic curve $E$ (the remnant of the dilaton of the
heterotic string). Further analysis of such examples should shed
new light on the interplay between Seiberg-Witten theory,
integrability, field theory and quantum geometry.

\section{Examples \label{Exa}}

\subsection{Holomorphic differentials on a punctured sphere}

If Riemann sphere has punctures at the points $\lambda_i$,
$i=1,\ldots,N$, then the canonical basis in the space $\Omega^1$
is:
\be
d\omega_i = \frac{(\lambda_i-\lambda_N)d\lambda}{(\lambda - \lambda_i)
(\lambda - \lambda_N)}, \ \ \ i=1,\ldots,N-1
\ee
We assumed that the $A_i$ cycles wrap around the points $\lambda_i$,
while their conjugated $B_i$ connect $\lambda_i$ with the reference
puncture $\lambda_N$. Multiplication algebra of $d\omega_i$'s
is defined modulo
\be
dW = d\log P_N(\lambda) = \frac{dP_N(\lambda)}{P_N(\lambda)},
\ee
$P_N(\lambda) = \prod_{i=1}^N (\lambda - \lambda_i)$, and
it is obviously associative.

The periods $a_i$ depend on the choice of the generating
differential $dS = \Lambda dW$. There are two essentially different
choices $\Lambda = \lambda$ and $\Lambda = \log\lambda$, i.e.
\be
dS^{(4)} = \lambda \ d\log P_N(\lambda) \ \ \ {\rm and} \ \ \
dS^{(5)} = \log\lambda\  d\log P_N(\lambda)
\ee
In order to fulfill the requirement (\ref{holreq}) one should
assume that $\sum_{i=1}^N \lambda_i = 0$ in the case of
$dS^{(4)}$, while $\prod_{i=1}^N \lambda_i = 1$ in the case
of $dS^{(5)}$. Since $A_i$ cycle just wraps around the point $\lambda
=\lambda_i$, the $A_i$-periods of such $dS$ are
\be
a_i^{(4)}=\oint_{\lambda_i}dS^{(4)}=\lambda_i,\\
a_i^{(5)}=\oint_{\lambda_i}dS^{(5)}=\log\lambda_i
\ee

The corresponding residue formulas are
\be
{\cal F}^{(4)}_{ijk} = \sum_{m=1}^N \stackreb{\lambda_m}{\res}
\frac{d\omega_id\omega_jd\omega_k}{d\lambda\ d\log P_N}, \nn \\
{\cal F}^{(5)}_{ijk} = \sum_{m=1}^N \stackreb{\lambda_m}{\res}
\lambda\frac{d\omega_id\omega_jd\omega_k}{d\lambda\ d\log P_N},
\ \ \ i,j,k=1,\ldots,N-1
\ee
and they both provide solutions to the WDVV equations \cite{MMM2}.
The prepotentials are:
\be
{\cal F}^{(4)}(a_i) = \frac{1}{2}\sum_{1\leq i < j \leq N}
(a_i - a_j)^2\log(a_i-a_j), \ \ \ \sum_{i=1}^N a_i = 0,
\ee
and
\be
{\cal F}^{(5)}(a_i) =\sum_{1\leq i < j \leq N}
\widetilde Li_3\left(e^{a_i-a_j}\right)-
{N\over 2}\sum_{1\leq i < j < k \leq N} a_i a_j a_k, \ \ \
\sum_{i=1}^N a_i = 0,\\
\partial^2_x \widetilde Li_3\left(e^x\right)\equiv \log 2\sinh x,\ \ \
\widetilde Li_3\left(e^x\right)={1\over 6} x^3 -{1\over 4}
Li_3\left(e^{-2x}\right)
\ee
They describe the perturbative
limit of the $N=2$ supersymmetric $SU(N)$ gauge models in $4d$ \cite{4dSW}
and $5d$ \cite{N} respectively.

If the punctures $\lambda_i$  are not all independent,
the same formulas provide solutions to the WDVV equations,
associated with the other simple groups: $SO(N)$,
$Sp(N)$, $F_4$ and $E_{6,7,8}$ ($G_2$ does not have enough
moduli to provide non-trivial solutions to the WDVV eqs).
If $P_N$ is substituted by
\be
P_N \rightarrow \frac{P_N}{Q_{N_f}^{1/2}} = \frac{\prod_{i=1}^N
(\lambda - \lambda_i)}{\prod_{\iota = 1}^{N_f} (\lambda - m_\iota)^{1/2}},
\ee
one gets solutions, interpreted as (perturbative limits of) the gauge models
with matter supermultiplets in the first fundamental representation.
Inclusion of matter in other representations seems to
destroy the WDVV equations, at least, generically; note that such models do
not arise in a natural way from string compactifications, and there
are no known curves associated with them in the Seiberg-Witten theory
(see \cite{MMM2} for details).

\subsection{Holomorphic differentials on hyperelliptic curves}

{\it Non-perturbative} deformations of the above prepotentials
arise when the punctures on Riemann sphere are blown up to form
handles of the hyperelliptic curve:
\be
w + \frac{1}{w} = 2\frac{P_N(\lambda)}{Q(\lambda)^{1/2}_{N_f}}, \nn \\
w - \frac{1}{w} = 2\frac{Y(\lambda)}{Q(\lambda)^{1/2}_{N_f}}, \nn \\
Y^2(\lambda) = P_N^2(\lambda) - Q_{N_f}(\lambda)
\ee
These curves, together with the corresponding differentials $dS$
\be
dS^{(4)} = \lambda\frac{dw}{w}, \ \ \
dS^{(5)} = \log\lambda \frac{dw}{w},
\ee
(i.e. $dW = \frac{dw}{w}$ and $d\Lambda^{(4)} = d\lambda$, $d\Lambda^{(5)}
= \frac{d\lambda}{\lambda}$) are implied by integrable models of the
Toda-chain family \cite{GKMMM,MW,NT,N}.
Together with the residue formula (\ref{resfdiff}) these provide the
solution to the WDVV equations. Further details about these
examples can be found in \cite{MMM,MMM2}.

\subsection{Other examples}

The very natural question is what happens with the WDVV equations
for Toda chain models, associated with the exceptional groups. The
problem is that the associated spectral curves are not hyperelliptic --
at least naively. Still they have enough symmetries to make our
general reasoning working, but this requires a special investigation.

The number of examples can be essentially increased by the study of
various integrable hierarchies, peculiar configurations of punctures
etc. In recent paper \cite{KriL} it was actually suggested that
-- at least in peculiar models -- $dS$ can be expressed through the
Baker-Akhiezer function: $dS = \Lambda d\log\Psi$. The study of such
examples involves generic expressions for the prepotentials, sometimes
with the higher time-variables included, see \cite{KriWhi,GKMMM,NT,IM2}.

Of more importance should be development in another direction.
We mentioned above that transition from $4d$ to $5d$ models \cite{N}
includes just the change of parametrization of punctured Riemann sphere:
from ``plane'' parametrization to the ``annulus'' one ($\lambda
\rightarrow \log\lambda$). The crucially interesting lift to $6d$
models requires interpretation of $\lambda$ as a coordinate on
elliptic curve. This is rather straightforward, and it is extremely
interesting to know, if this transition breaks the WDVV equations
(as happens in case of elliptization of the dual variable $w$).
See \cite{GMMM} for preliminary discussion of the relevant
elliptic $XYZ$ model.

The main goal of these studies -- as mentioned in the Introduction --
can be better understanding of quantum cohomologies and structures
behind the prepotentials of the string models.

\section{Acknowledgments}

We are indebted for illuminating discussion to many people,
especially to  E.Akhmedov, M.Alishahiha,
F.Ardalan, G.Curio, J.de Boer, B.de Wit, B.Dubrovin, A.Gorsky,
I.Krichever, A.Levin, A.Losev, D.L{\"u}st, Yu.Manin, M.Matone,
N.Nekrasov and A.Rosly.

This work was partially supported by the grants
RFFI 96-01-01106, INTAS 93-0633 (A.Marshakov),
RFFI 96-01-00887, INTAS-RFBR-95-0690 (A.Mironov)
and RFFI 96-15-96939 (A.Morozov).
A.Morozov acknowledges the support of DFG and
the hospitality of Humboldt University,
Berlin and IFH, Zeuthen.


\begin{thebibliography}{12}

\bibitem{W}
E.Witten, Surv.Diff.Geom. {\bf 1} (1991) 243.
\bibitem{DVV}
R.Dijkgraaf, E.Verlinde and H.Verlinde, Nucl.Phys. {\bf B352} (1991) 59.
\bibitem{D1}
B.Dubrovin, {\sl Geometry of 2D topological field theories},
hepth/9407018.
\bibitem{chr}
W.Lerche, C.Vafa and N.Warner, Nucl.Phys. {\bf B324} (1989) 427.
\bibitem{ref}
P.Griffits and J.Harris, {\sl Principles of Algebraic Geometry}, 1978.
\bibitem{KriWhi}
I.Krichever, {\sl The tau-function of the universal Whitham hierarchy, matrix
models and topological field theories}, Preprint LPTENS-92-18;
Comm.Pure Appl.Math. {\bf 47} (1994) 437.
\bibitem{MMM}
A.Marshakov, A.Mironov and A.Morozov, hepth/9607109, to appear in
Phys.Lett. {\bf B}.
\bibitem{MMM2} A.Marshakov, A.Mironov and A.Morozov, preprint
FIAN/TD-15/96, ITEP/TH-46/96.
\bibitem{SW}
N.Seiberg and E.Witten, Nucl.Phys. {\bf B426} (1994) 19; {\bf B431} (1994) 484.
\bibitem{M}
Yu.Manin, {\sl Frobenius manifolds, quantum cohomology and moduli spaces},
Preprint MPI, 1996.
\bibitem{KoMa}
M.Kontsevich and Yu.Manin, Comm.Math.Phys. {\bf 164} (1994) 525.
\bibitem{P}
L.Caporaso and J.Harris, alg-geom/9608025
\bibitem{G}
E.Getzler, alg-geom/9612004.
\bibitem{BM}
G.Bonelli and M.Matone, hepth/9605090.
\bibitem{BW} J.de Boer and B.de Wit, private communication.
\bibitem{Whith}
I.Krichever, Func.An. \& Apps. {\bf 22} (1988), n 3, 37;\\
B.Dubrovin and S.Novikov, Uspekhi Mat.Nauk, {\bf 44} (1989) N6, 29;\\
B.Dubrovin, Comm.Math.Phys. {\bf 145} (1992) 195.
\bibitem{Hit}
B.Dubrovin, I.Krichever and S.Novikov, in
{\sl Sovremennye problemy matematiki (VINITI), Dynamical systems - 4}
(1985) 179;\\
N.Hitchin, Duke.Math.Jour. {\bf 54} (1987) 91;\\
E.Markman, Comp.Math. {\bf 93} (1994) 255;\\
A.Gorsky and N.Nekrasov, hepth/9401021;\\
R.Donagi and E.Witten, hepth/9510101.
\bibitem{IM2}
H.Itoyama and A.Morozov, hepth/9512161.
\bibitem{4dSW}
A.Klemm, W.Lerche, S.Theisen and S.Yankelovich, hepth/9411048;
Phys.Lett., {\bf B344} (1995) 169;\\
P.Argyres and A.Farragi, hepth/9411057;
{\sl Phys.Rev.Lett.,} {\bf 74} (1995) 3931.
\bibitem{N}
N.Nekrasov, preprint ITEP/TH-26/96, HUTP-96/A023; hepth/9609219.
\bibitem{GKMMM}
A.Gorsky, I.Krichever, A.Marshakov, A.Mironov and A.Morozov,
Phys.Lett., {\bf B355} (1995) 466; hepth/9505035.
\bibitem{MW}
E.Martinec and N.Warner, hepth/9509161.
\bibitem{NT}
T.Nakatsu and K.Takasaki, hepth/9509162.
\bibitem{KriL}
I.Krichever, hep-th/9611158.
\bibitem{GMMM}
A.Gorsky, A.Marshakov, A.Mironov and A.Morozov,
in {\it Problems in Modern Theoretical Physics},
Dubna 1996, 44; hepth/9604078.

\end{thebibliography}
\end{document}